# Modified Steinberg–Guinan elasticity model to describe softening-hardening dual anomaly in vanadium

Hao Wang[1], Yuan-Chao Gan[1], Xiang-Rong Chen[2], Yi-Xian Wang[3], Hua Y. Geng[1, 4*]

[1] *National Key Laboratory of Shock Wave and Detonation Physics, Institute of Fluid Physics, CAEP, Mianyang 621900, People's Republic of China*

[2] *College of Physics, Sichuan University, Chengdu 610065, People's Republic of China*

[3] *College of Science, Xi'an University of Science and Technology, Xi'an 710054, People's Republic of China*

[4] *HEDPS, Center for Applied Physics and Technology, and College of Engineering, Peking University, Beijing 100871, People's Republic of China*

**Abstract:** Constitutive models are essential for describing the mechanical behavior of materials under high temperatures and pressures, among which the Steinberg–Guinan (SG) model is widely adopted. Recent work has discovered a peculiar dual anomaly of compression-induced softening and heating-induced hardening in the elasticity of compressed vanadium [Phys. Rev. B **104**, 134102 (2021)], which is beyond the capability of the SG model to describe. In this work, a modified SG constitutive model is proposed to embody such an anomalous behavior. Elemental vanadium is considered as an example to demonstrate the effectiveness of this improved model in describing the dual anomalies of mechanical elasticity. This new variant of the SG model can also be applied to other materials that present an irregular variation in the mechanical elasticity, and is important to faithfully model and simulate the mechanical response of materials under extreme conditions.

## 1. Introduction

Universal constitutive models valid over a wide pressure–temperature range are crucial for describing the strengths of materials subjected to dynamic loading. The Steinberg–Guinan (SG) constitutive model[1] decouples the influences of temperature and pressure on the elasticity and strength of materials. Hence, it has wide applicability, which includes the processing of metals for shock or impact engineering applications. A variant of the SG model has been proposed to improve its performance in the solid–liquid mixed zone[2]. Notably, SG and its offspring models are built upon the basic assumptions that a pressure should induce material hardening and temperature should cause material softening.

---

* *Corresponding author. E-mail:* s102genghy@caep.cn

However, the elastic properties of metals under extreme conditions are more complex than previously thought. For example, using theoretical calculations, Landa et al.[3] predicted that the shear elastic constant $C_{44}$ of BCC vanadium might decrease with pressure and become negative at ~110 GPa, thus exhibiting the counter-intuitive phenomenon of compression-induced softening (CIS). Koči et al.[4] also reported similar anomalous CIS behavior in vanadium. Wang et al.[5,6] elaborated on the origin of this phenomenon and found that both vanadium and niobium exhibit a closely related heating-induced hardening (HIH) behavior that mainly arises from the thermo-electron effect. A recent study[7] combining theoretical analyses and experimental measurements provided direct evidence for the existence of this softening-hardening (i.e., CISHIH) dual anomaly in compressed vanadium. It was also revealed that along the shock Hugoniot, the SG model and its offspring variants failed to describe completely the variation in the elasticity and sound velocity of vanadium.

Such irregular variations in elasticity are not limited to vanadium and niobium[3-6]; compressed tantalum[4,8] and were also predicted to exhibit similar behavior, as were other materials[9-12]. However, the temperature term in the SG model describes only the thermal softening effect caused by lattice vibrations[1]; Walker[13,14] showed that this lattice thermal softening in vanadium and niobium occurs only when $T > T_{max}$ (with $T_{max}$ ~0.74 $T_{melt}$). Rudd et al.[15] amended the SG model to account for the anomalous softening, but their work specifically focused on interpolating the ab initio data with the influence of the BCC→RH distortion of vanadium[16,17] at high pressures. No attempt was made to include the HIH effect. Therefore, a genuine constitutive model that can model the CISHIH dual anomaly is still lacking.

In this study, we extend the SG model to consider CISHIH behavior in the elasticity of materials at high temperatures and pressures. Compressed vanadium is employed as an example, and the model's parameters are determined using finite-temperature density functional theory (DFT) calculations. We show that this modified SG model accurately describes the mechanical dual anomaly in the BCC phase of vanadium.

## 2. Theory and computational details

### 2.1 Steinberg–Guinan model

The SG model approximates the variation in the shear modulus as a function of the pressure and temperature as follows[1].

$$G(P(\eta),T) = G_0[1 + AP\eta^{-1/3} - D(T-300)] \quad (1)$$

$$A = \frac{1}{G_0}\frac{dG}{dP}\bigg|_{P=P_0, T=T_0} \quad D = -\frac{1}{G_0}\frac{dG}{dT}\bigg|_{P=P_0, T=T_0} \quad (2)$$

where $G_0$ is the ambient shear modulus, $\eta = V_0/V$ is the volume compression ratio, and $A$ and $D$ represent the rate of relative changes in the shear modulus of the material with pressure and temperature, respectively. Apart from the $\eta^{-1/3}$ factor, Eq. (1) is actually a linear expansion of $G(P, T)$ near the reference point of (0 GPa, 300 K). The contribution of the $\eta^{-1/3}$ factor to the pressure dependence is nonlinear, yet it remains monotonically increasing. In principle, the SG model should be valid only in the vicinity of the expansion point, and any complex variations, such as a non-monotonic variation in pressure or nonlinear variation in temperature[3, 5, 7, 16], are beyond its applicable realm.

### 2.2 Modified Steinberg–Guinan model

The apparent limitation of the SG model motivates us to modify it to expand its applicability for irregular variations in the shear modulus, particularly the CISHIH dual anomaly. From previous investigations, it is clear that the CISHIH anomaly and other irregular variations occur only in specific temperature and pressure ranges. It is thus rational to assume that outside these ranges, the shear modulus should reduce to that of the original SG model. We consider this as an asymptotic condition that must be satisfied when constructing the improved model. Namely, the correction to the SG model only considers effects within the range where an irregular variation such as the CISHIH dual anomaly is prominent. This precise strategy preserves the framework integrity of the SG model as much as possible.

By thoroughly inspecting the shape of the CISHIH dual anomaly and other irregular variations of the shear modulus in metals[5-7], we discovered that their characteristic is similar to a peak function in the pressure and temperature space. Inspired by this observation, we propose a modification to the SG model to incorporate anomalous variations in the shear modulus. The expressions of the modified model are as follows.

$$G(P(\eta),T) = G_0[1 + AP(\eta)\eta^{-1/3} - D(T-300)] + G_e(\eta,T) \qquad (3)$$

$$G_e(\eta,T) = y_0(T) + H(T) Exp\left(-0.5\left(\frac{\eta - x_c(T)}{\omega(T)}\right)^2\right) \qquad (4)$$

where term $G_e$ bears all amendments to the SG model; $y_0$, $H$, $x_c$, and $\omega$ are coefficients depending only on the temperature, and are expressed as follows:

$$\begin{aligned} y_0(T) &= c_0 + c_1 T + c_2 T^2 \\ x_c(T) &= x_0 + x_1 T \\ H(T) &= h_0 + h_1 T \\ \omega &= \begin{cases} \alpha_0 + \alpha_1 T + \alpha_2 T^2, & \text{if } \eta \leq x_c \\ \beta_0 + \beta_1 T + \beta_2 T^2, & \text{if } \eta \geq x_c \end{cases} \end{aligned} \qquad (5)$$

The exponential term in $G_e$ decays to zero when far away from $x_c$ or with a large enough $\omega$, thus satisfying the asymptotic condition in the pressure–temperature (P–T) space. Term $y_0$ is introduced to compensate for the variation in the shear modulus with temperature outside the range with irregular behavior.

It should be noted that in Eq. (3), we expressed pressure as a function of $\eta$ explicitly. That is, we transformed the shear modulus from the P–T to the $\eta$–T space, and all corrections to the SG model were performed in the latter space. This is possible by employing the third-order Birch–Murnaghan equation of state (EOS) to relate the pressure and volume compressibility.

$$P(V) = \frac{3B_0}{2}\left[\eta^{\frac{7}{3}} - \eta^{\frac{5}{3}}\right]\left\{1 + \frac{3}{4}\left(B_0' - 4\right)\left[\eta^{\frac{2}{3}} - 1\right]\right\} \qquad (6)$$

Eqs. (3–6) describe the modified SG model.

### 2.3 Computational details

In this work, DFT calculations were performed using VASP [18, 19]. A plane-wave basis set was employed with a kinetic energy cut-off of 900 eV. The electron-core interaction was described by a projector-augmented wave pseudopotential [20], which contains 13 valence electrons ($3s^23p^63d^34s^2$) for vanadium. The electron exchange correlation was handled by a generalized gradient approximation with the Perdew–Burke–Ernzerhof parameterization[21]. The k-point sampling in the irreducible Brillouin zone was performed with a resolution of $2\pi \times 0.05$ Å$^{-1}$. In the self-consistent field calculation, the energy (force) convergence criterion was set to $10^{-8}$ eV/atom (0.001 eV/Å). The thermal electronic contribution was fully considered via the electronic free-energy functional of the finite-temperature DFT of Mermin [22]. The elastic constants of single-crystal vanadium were obtained using the energy-strain method [23], which was implemented in the MyElas software program [24]. The shear modulus of the polycrystalline sample was derived using the Voigt–Reuss–Hill method [25, 26]:

$$G_V = \frac{C_{11} - C_{12} + 3C_{44}}{5} \quad (7)$$

$$G_R = \frac{5C_{44}(C_{11} - C_{12})}{4C_{44} + 3(C_{11} - C_{12})} \quad (8)$$

$$G_H = \frac{G_V + G_R}{2} \quad (9)$$

### 3. Results and discussion

To determine the parameters of the modified SG (MSG elasticity) model, theoretical or experimental data of the shear modulus at different pressures and temperatures are required. Herein, we employ DFT data as an input to evaluate the ability of the MSG elasticity model to describe the irregular variations in the elasticity. The monocrystalline $C_{44}$ and derived polycrystalline shear modulus of vanadium calculated by DFT are displayed in Fig. 1. The CISHIH dual anomaly in them is evident. This non-monotonic variation in the shear modulus $G$ with pressure as well as its increase with temperature cannot be described by the original SG model[1], or by other variants. Parameterization of the MSG elasticity model involves five steps:

(**I**) Determine the third-order Birch–Murnaghan EOS by fitting it to the cold compression curve, as shown in the inset of Fig. 2(a). This fixes the values of $B_0$ and $B_0$', as well as the function $P(\eta)$.

(**II**) By inspecting the DFT data of the shear modulus, identify the pressure range where CIS is prominent.

(**III**) Using $P(\eta)$ and the shear modulus data at zero temperature outside the CIS range, re-parameterize the compression part of the SG model, and fix the values of $G_0$ and $A$, as shown in Fig. 2(a).

(**IV**) Taking the compression of the SG model (i.e., $G(P) = G_0(1+AP\eta^{-1/3})$) as the reference, parameterize the $G_e$ part of the MSG elasticity model by fitting to the DFT-calculated shear modulus at different pressures and temperatures, as shown in Fig. 2(b). This fixes the values of $y_0$, $x_c$, $H$, $\omega_1$, and $\omega_2$.

(**V**) The lattice thermal softening term $D(T - 300)$ is taken from the empirical value of the original SG model[15].

In this manner, we parameterized the MSG elasticity model, and all parameters were determined for vanadium as listed in Table I.

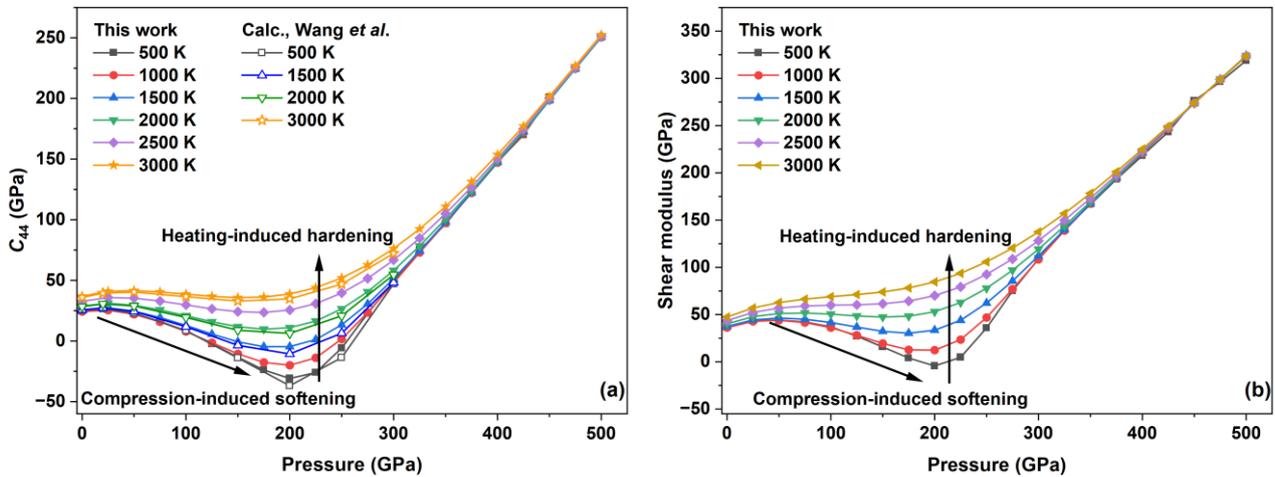

**Fig. 1**. (Color online) DFT-calculated $C_{44}$ and shear modulus $G$ of vanadium at various pressures and temperatures, in which the CISHIH dual anomaly is evident. The theoretical data of Wang *et al.*[5] are shown for comparison in (a).

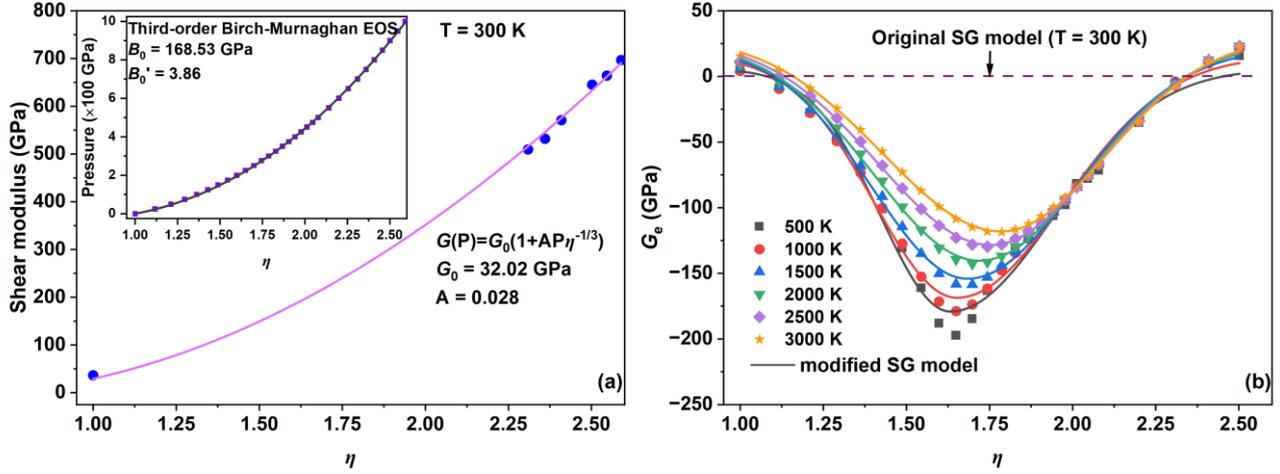

**Fig. 2.** (Color online) (a) Re-parameterizing the compression part of the SG model (solid line) by using the cold curve of the DFT data (solid points). The inset shows the $P(\eta)$ function (solid line) of the third-order Birch–Murnaghan EOS against the DFT data (solid points). (b) Parameterizing the $G_e$ part of the MSG elasticity model by fitting to the DFT-calculated shear modulus at various pressures and temperatures. The solid points represent the DFT data, and the solid lines represent the MSG elasticity model. The original SG model at 300 K is taken as the reference and also plotted for comparison.

**Table I**. Parameters of the modified SG model for compressed vanadium in the BCC phase.

| $V_0$ (Å³/atom) | $B_0$ (GPa) | $B_0'$ | $G_0$ (GPa) | $A$ (GPa⁻¹) | $D$ (K⁻¹) | $c_0$ (GPa) |
|---|---|---|---|---|---|---|
| 13.47 | 168.53 | 3.86 | 32.02 | 0.028 | 2.06×10⁻⁴ | -2.801 |
| $c_1$ (GPa/K) | $c_2$ (GPa/K²) | $x_0$ | $x_1$ (K⁻¹) | $h_0$ (GPa) | $h_1$ (GPa/K) | $\alpha_0$ |
| 0.021 | -3.262×10⁻⁶ | 1.590 | 6.514×10⁻⁵ | -199.538 | 0.0167 | 0.157 |
| $\alpha_1$ (1/K) | $\alpha_2$ (1/K²) | $\beta_0$ | $\beta_1$ (1/K) | $\beta_2$ (1/K²) | | |
| 1.073×10⁻⁴ | -1.396×10⁻⁸ | 0.305 | 1.905×10⁻⁵ | -5.063×10⁻⁹ | | |

The improvement of the MSG elasticity model is demonstrated in Fig. 3 by comparison with the original SG model for vanadium over a wide P–T range. Here, we do not consider the possible BCC-RH-BCC transition[5, 17, 27, 28], but rather focus on the description capability of the SG and MSG elasticity models when irregular variations such as CISHIH dual anomalies exist. In Fig. 3, the melting curve is taken from Ref. [29]. It is evident from Fig. 3(a) that the original SG model demonstrates only monotonical increases in the shear modulus of vanadium with pressure, and a linear decrease with temperature. This behavior is contradictory to what is observed in experiment [7] and predicted by ab initio calculations [3,5,16]; thus, it is qualitatively wrong.

By contrast, as shown in Fig. 3(b), the MSG elasticity model describes the CIS in the shear modulus $G$ for BCC vanadium at a low temperature when the pressure is less than 200 GPa. At higher pressures, $G$ monotonically increases with pressure. This behavior is in agreement with the current understanding of compressed vanadium. Furthermore, the HIH in $G$ is also successfully reproduced, persisting up to the melting temperature within the pressure range of 0–370 GPa. It is noteworthy that this hardening behavior in vanadium is weakened gradually when the pressure approaches 370 GPa. Beyond 370 GPa, compressed vanadium exhibits the usually expected heating-induced softening (HIS) in $G$, and this thermal softening is stronger in the vicinity of the melting curve. The division of the P–T space of solid vanadium into two separate regions, namely the HIH and HIS regions in $G$ as shown in Fig. 3(b), is a new understanding that the MSG elasticity model affords, which is not apparent from the DFT data displayed in Fig. 1(b). The MSG elasticity model also allows the deduction that at some temperature, the shear modulus of vanadium along the isotherm will have a flat region (i.e., $dG/dP = 0$) at low pressures.

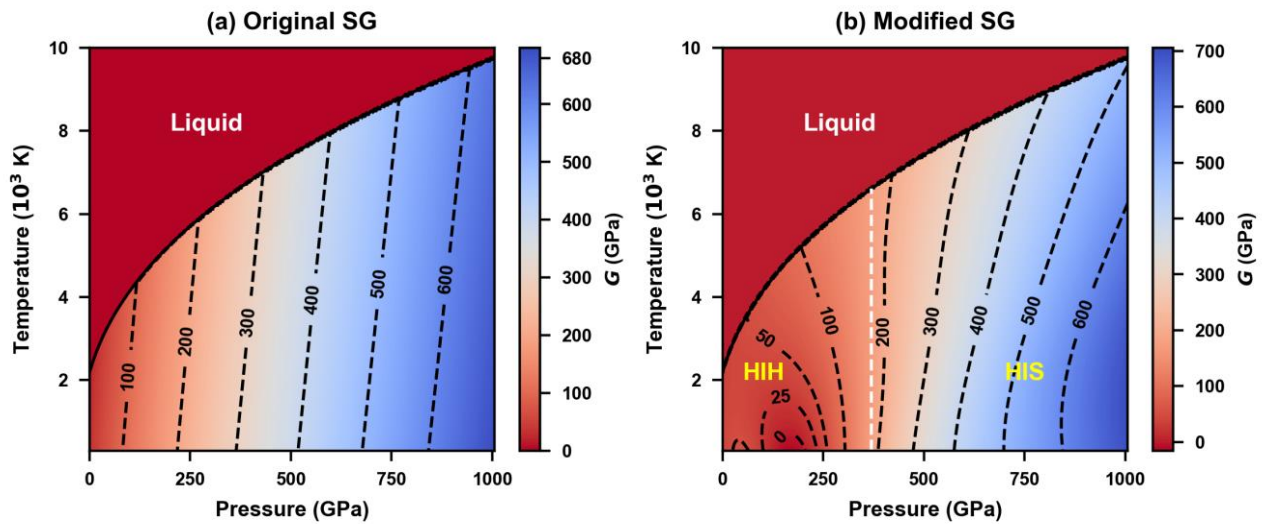

**Fig. 3.** (Color online) Contour lines of shear modulus of BCC vanadium at high temperatures and pressures as predicted by (a) the original SG model and (b) the MSG elasticity model. The shear

modulus contour values are indicated on the color maps. The melting curve of vanadium is taken from the experimental results[29].

In previous studies, most experimental data on elasticity were measured in terms of the sound velocity by the shock-wave method. To obtain calculated shear sound velocities and invert the longitudinal sound velocity by the reported bulk sound velocity[7, 30], we incorporate the Hugoniot equation of state data directly into the modified SG elasticity model. The data points for the Hugoniot equation of state H (P, V, T) are extracted from the experimental[31, 32] and computational values reported[29, 33]. The results are compared to the experimental data [7, 30, 34], and the prediction given by the original SG model[1] in Fig. 4. Compared with the original SG model, the MSG elasticity model is closer to the experimental. In particular, the low-pressure flat region and the monotonically increasing region at pressures beyond 100 GPa are well reproduced, whereas the original SG model indicates a featureless monotonic increment in the shear sound velocity. In the high-pressure zone, the sound velocity values of the MSG elasticity model are lower than those in the experimental data. This discrepancy may originate from overestimation of the effect of lattice softening on the sound speed by $-D$(T – 300). In future iterations, we plan to obtain the variations in shear modulus caused by lattice vibrational effects under different pressures and temperatures through DFT calculations, thereby providing a new simple expression to replace $-D$(T – 300) in the MSG elasticity model.

Moreover, for some materials, the DFT-calculated values might deviate from the experimental values; for example, the shear modulus of vanadium single crystals is lower than the experimental value, but the trend is consistent. At present, our fitting parameters are derived from the DFT data. In future work, we may utilize the discrepancy between the MSG elasticity model and the experimental values, $\Delta$ ($G_{\text{expt.}}$ - $G_{\text{MSG elasticity}}$), as the loss function during the fitting process, thereby increasing the correlation coefficient and consequently improving the alignment between the model and experimental values. Our primary focus in this research is the introduction of a novel model; the optimization algorithm for parameters is not a significant concern at this time. In addition, at higher temperature, the material enters the solid-liquid coexistence region or even the melting region, causing a dramatic decrease in the actual shear modulus, eventually reaching 0. This phenomenon cannot be accurately described by either the original SCG model or our modified model, and requires a different form of modification[2].

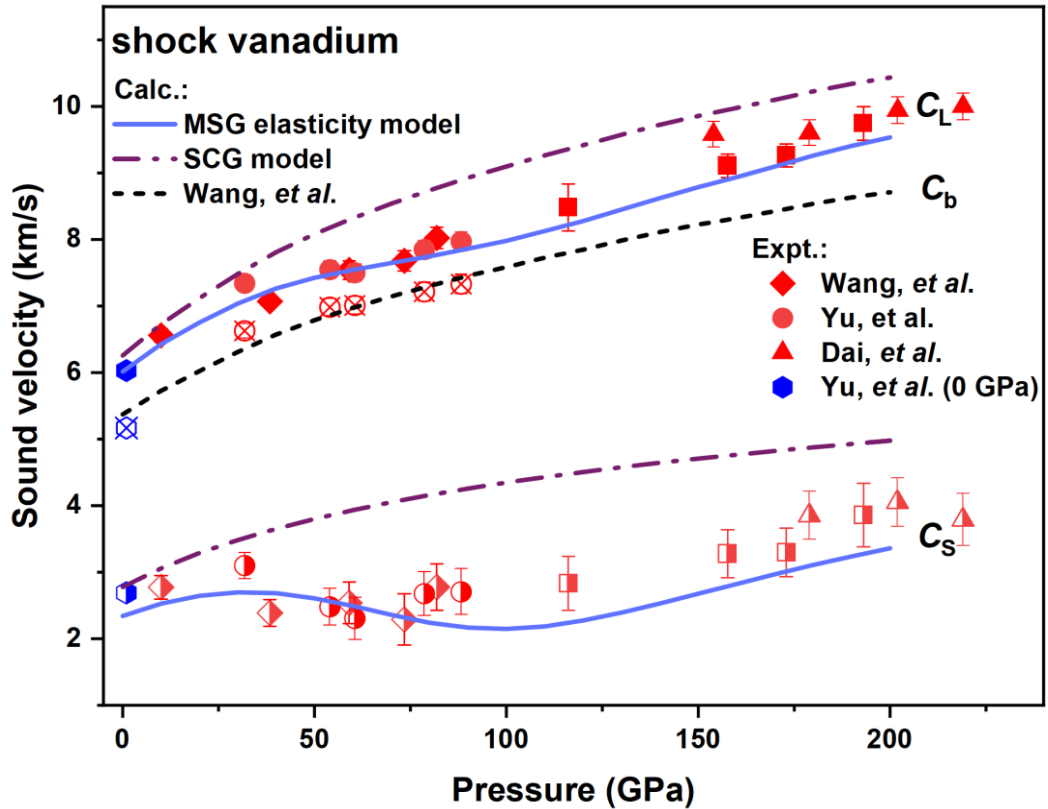

**Fig. 4**. (Color online) Comparison of the predicted sound velocity along the Hugoniot for polycrystalline vanadium calculated by the SG model[1], MSG elasticity model, and the shock experimental data of Dai [34], Yu [30], and Wang [7]. The reference bulk sound velocity data were obtained from Yu's experimental measurements[30] and Wang's theoretical calculations[7].

Using vanadium as an example, we have confirmed the robustness of the MSG elasticity model. It is feasible to apply the MSG elasticity model to other materials with similar mechanical behavior. However, note that while different materials may exhibit similar softening and hardening behaviors, the associated mechanisms might vary. Therefore, when experimental data are scarce, careful consideration should be given to obtaining reliable DFT data for fitting the correlation coefficient. The use of *ab initio* molecular dynamics might be a more generalized approach; however, its computational cost is extremely high. Machine learning techniques have effectively addressed this challenge. Through an active training approach, one can construct a machine learning potential that relies on fewer DFT calculations to carry out high-precision molecular dynamics simulations and establish a fitting dataset[35, 36], representing a potentially viable solution.

## 4. Conclusions

To expand the applicable range of the SG model, the MSG elasticity model was proposed. Using vanadium as a representative example, we demonstrated that the MSG elasticity model successfully captures the CISHIH dual anomaly in the shear modulus occurring at high pressures and temperatures. This improvement in modeling not only results in a more accurate strength model, but also leads to a new understanding regarding compressed vanadium (e.g., the division of the P–T space into HIH and HIS regions, and the flat range in $G$ at low pressures). This work is the first attempt to model the irregular shear modulus within the framework of the SG model, and the results suggest that this direction is promising. Further refinement of the MSG elasticity model could improve its quantitative agreement with experimental or ab initio data. The application of this modeling approach to other elastic moduli[7] and materials[4, 6, 8] is straightforward, and might be helpful for modeling and simulating the complex mechanical response of materials under extreme conditions.

## ACKNOWLEDGMENTS

This work was supported by the National Key R&D Program of China under Grant No. 2021YFB3802300, Science and Technology Foundation of State Key Laboratory of Shock Wave under, the NSAF under Grant No. U1730248 and U1830101 and the National Natural Science Foundation of China under Grant Nos. 11672274, 11872056, and 11904282. The simulation was performed using the resources provided by the Center for Comput. Mater. Sci. Software (CCMS, Tohoku University, Japan).

## AUTHOR DECLARATIONS
### Conflict of Interest

The authors have no conflicts to disclose.

## Author Contributions

**Hao Wang**: Methodology (equal); Software (lead); Visualization (lead); Writing – original draft (lead); Writing – review & editing (equal). **Yuan-Chao Gan**: Methodology (equal). **Xiang-Rong Chen**: Funding acquisition (equal); Methodology (equal); Writing – review & editing (equal). **Yi-Xian Wang**: Funding acquisition (equal); Writing – review & editing (equal). **Hua Y. Geng**: Idea conceiving; Funding acquisition (equal); Methodology (equal); Writing – review & editing (lead); Project administration (lead).

## DATA AVAILABILITY

The data that support the findings of this study are available from the corresponding author upon reasonable request.

## References

1. D. J. Steinberg, S. G. Cochran and M. W. Guinan, J. Appl. Phys. **51** (3), 1498-1504 (1980).
2. X. Yang, X. Zeng, F. Wang, H. Zhao, J. Chen and Y. Wang, Mech. Mater. **155**, 103775 (2021).
3. A. Landa, J. Klepeis, P. Söderlind, I. Naumov, O. Velikokhatnyi, L. Vitos and A. Ruban, J. Phys. Chem. Solids **67** (9-10), 2056-2064 (2006).
4. L. Koči, Y. Ma, A. R. Oganov, P. Souvatzis and R. Ahuja, Phys. Rev. B **77** (21), 214101 (2008).
5. Y. X. Wang, Q. Wu, X. R. Chen and H. Y. Geng, Sci. Rep. **6**, 32419 (2016).
6. Y. X. Wang, H. Y. Geng, Q. Wu, X. R. Chen and Y. Sun, J. Appl. Phys. **122** (23), 235903 (2017).
7. H. Wang, J. Li, X. M. Zhou, Y. Tan, L. Hao, Y. Y. Yu, C. D. Dai, K. Jin, Q. Wu, Q. M. Jing, X. R. Chen, X. Z. Yan, Y. X. Wang and H. Y. Geng, Phys. Rev. B **104** (13), 134102 (2021).
8. Y. Zhang, C. Yang, A. Alatas, A. H. Said, N. P. Salke, J. Hong and J.-F. Lin, Phys. Rev. B **100** (7), 075145 (2019).
9. S. Ağduk and G. Gökoğlu, J. Alloys Compd. **520**, 93-97 (2012).
10. Z. Wu, J. F. Justo and R. M. Wentzcovitch, Phys. Rev. Lett. **110** (22), 228501 (2013).
11. Y. O. Kvashnin, W. Sun, I. Di Marco and O. Eriksson, Phys. Rev. B **92** (13), 134422 (2015).

12. P. V. Sreenivasa Reddy, V. Kanchana, G. Vaitheeswaran, P. Modak and A. K. Verma, J. Appl. Phys. **119** (7), 075901 (2016).

13. E. Walker, Solid State Commun. **28**, 587-589 (1978).

14. Y. Talmor, E. Walker and S. Steinemann, Solid State Commun. **23**, 649 (1977).

15. R. E. Rudd and J. E. Klepeis, J. Appl. Phys. **104** (9), 093528 (2008).

16. B. Lee, R. E. Rudd, J. E. Klepeis and R. Becker, Phys. Rev. B **77** (13), 134105 (2008).

17. Y. Ding, R. Ahuja, J. Shu, P. Chow, W. Luo and H. K. Mao, Phys. Rev. Lett. **98** (8), 085502 (2007).

18. G. Kresse and J. Furthmuller, Comput. Mater. Sci. **6**, 15-50 (1996).

19. G. Kresse, Phys. Rev. B **54** (16), 11169-11186 (1996).

20. P. E. Blöchl, Phys. Rev. B **50** (24), 17953-17979 (1994).

21. J. P. Perdew, K. Burke and M. Ernzerhof, Phys. Rev. Lett. **77** (18), 3865-3868 (1996).

22. N. D. Mermin, Phys. Rev. **137** (5A), A1441-A1443 (1965).

23. Z. J. Wu, E. J. Zhao, H. P. Xiang, X. F. Hao, X. J. Liu and J. Meng, Phys. Rev. B **76** (5), 054155 (2007).

24. H. Wang, Y. C. Gan, H. Y. Geng and X.-R. Chen, Comput. Phys. Commun. **281**, 108495 (2022).

25. R. Hill, Proc. Phys. Soc. A **65**, 349-354 (1952).

26. A. Reuss, Ztschr.f.angew.Math.und Mech. **8**, 49-58 (1929).

27. M. G. Stevenson, E. J. Pace, C. V. Storm, S. E. Finnegan, G. Garbarino, C. W. Wilson, D. McGonegle, S. G. Macleod and M. I. McMahon, Phys. Rev. B **103** (13), 134103 (2021).

28. Y. Akahama, S. Kawaguchi, N. Hirao and Y. Ohishi, J. Appl. Phys. **129** (13), 135902 (2021).

29. Y. Zhang, Y. Tan, H. Y. Geng, N. P. Salke, Z. Gao, J. Li, T. Sekine, Q. Wang, E. Greenberg, V. B. Prakapenka and J.-F. Lin, Phys. Rev. B **102** (21), 214104 (2020).

30. Y. Yu, Y. Tan, C. Dai, X. Li, Y. Li, Q. Wu and H. Tan, Appl. Phys. Lett. **105** (20), 201910 (2014).

31. J. M. Foster, A. J. Comley, G. S. Case, P. Avraam, S. D. Rothman, A. Higginbotham, E. K. R. Floyd, E. T. Gumbrell, J. J. D. Luis, D. McGonegle, N. T. Park, L. J. Peacock, C. P. Poulter, M. J. Suggit and J. S. Wark, J. Appl. Phys. **122** (2), 025117 (2017).

32. R. G. Mcqueen and S. P. Marsh, J. Appl. Phys. **31** (7), 1253-1269 (1960).

33. P. F. Weck, P. E. Kalita, T. Ao, S. D. Crockett, S. Root and K. R. Cochrane, Phys. Rev. B **102** (18), 184109 (2020).

34. C. Dai, X. Jin, X. Zhou, J. Liu and J. Hu, J. Phys. D: Appl. Phys. **34**, 3064-3070 (2001).

35. I. S. Novikov, K. Gubaev, E. V. Podryabinkin and A. V. Shapeev, Mach. Learn.: Sci. Technol. **2** (2), 025002 (2021).

36. Y. Zhang, H. Wang, W. Chen, J. Zeng, L. Zhang, H. Wang and W. E, Comput. Phys. Commun. **253**, 107206 (2020).